\documentclass[12pt]{iopart}

\usepackage{iopams}
\usepackage{graphicx}
\begin{document}

\title{Massive black hole binaries in gas-rich galaxy mergers; multiple regimes of orbital
decay and interplay with gas inflows}

\author{Lucio Mayer}

\address{Institute for Theoretical Physics,
University of Z\"urich,
Winterthurestrasse 190, 8057 Z\"urich, Switzerland, and

Kavli Institute for Theoretical Physics, UC Santa Barbara, 552 University Road Santa Barbara, CA 93106, United States}
\ead{lmayer@physik.uzh.ch}

\begin{abstract}

We revisit the phases of the pairing and sinking of BHs in galaxy mergers and circunmunclear disks in light of the results 
of recent simulations with massive BHs embedded in predominantly gaseous backgrounds. After a general overview we discuss 
the importance of a fast orbital decay regime
dominated by global disk torques rather than by the local dynamical friction wake. This regime can dominate at BH binary separations
of a few tens of parsecs and below, following a phase of orbital circularization dominated by local dynamical friction. It is
similar to Type-I migration in planetary evolution. It can bring the black holes to
separations small enough for gravitational waves to take over on a timescale ranging from less than $\sim 10^7$ yr to up to $10^8$ yr, 
depending on whether the interstellar medium is smooth or clumpy.
Eventual gap opening at sub-pc scale separations slows down but does not interrupt the orbital decay.Subsequently, we discuss a new intriguing
connection between the conditions required for rapid orbital decay of massive BH binaries and those required for prominent gas inflows
in gas-rich galaxies. We derive a condition for the maximum inflow rate that a circumnuclear disk can host while still maintaining a 
sufficiently high gas density at large radii to sustain the decay of a BH binary. We find that gas inflows rates exceeding 10 $M_{\odot}$/yr, 
postulated to form massive BH seeds in some direct collapse models, would stifle the sinking of massive BH binaries in gas-dominated galactic 
nuclei. Vice-versa, lower inflow rates, below a solar mass per year, as required to feed typical AGNs, are compatible with a fast orbital decay 
of BH binaries across a wide range of masses.

\end{abstract}

\section{Introduction}

The evolution of massive black hole (BH) binaries in gaseous backgrounds has become an active field of research in the last decade.
Galactic nuclei hosting a significant gaseous component are expected to be ubiquitous in the modern scenario of hierarchical galaxy
formation since not only they are present in disk-dominated galaxies, which host relatively light central black holes 
(Greene al. 2010), but are also
expected in the disky high redshift progenitors of present-day early-type galaxies (Feldmann et al. 2010), 
the hosts of the most prominent among present-day supermassive black holes (SMBHs) (Kormendy \& Richstone 1995; Ferrarese \& Merritt 2000).
When these massive BHs merge they should produce one of the loudest gravitational waves signal, that can be
detected with future gravitational wave experiments even at cosmological distances (Vecchio 2004). 
Most of the effort of theorists so far has focused on understanding the evolution of the binaries when the BH dynamics can still 
be treated in the newtonian regime,
although there have been isolated studies of late stages of the decay of BH binaries using a pseudonewtonian
potential (Bogdanovic et al. 2008). Indeed the evolutionary phase preceding the stage at 
which gravitational wave emission becomes important is still not fully understood. This is a 
pressing issue now that numerical relativity
simulations have progressed far enough to be able to follow the final 
coalescence of massive BHs (Baker et al. 2006; Rezzolla 2009).
It is thus fair to say that we still do not know the timescales and efficiency of the BH pairing and merging process due to our
incomplete knowledge of the newtonian phase, despite the fact that semi-analytical models (Volonteri et al. 2003) and cosmological
simulations (di Matteo et al. 2008) assume that massive BHs merge instantaneously when their host galaxies merge.
In this paper we will focus on recent 
developments concerning BH binary decay in the netwonian regime. This still requires modeling hydrodynamics, 
gravity and radiation physics
across several orders of magnitude in density, from galactic scales to sub-nuclear scales when the separation
of the holes decreases below parsecs, a daunting task for even the most advanced astrophysical codes
available. We will also outline a possible interesting interplay between the orbital decay of massive BH binaries 
and the formation of  massive BH seeds via direct gas collapse in galactic nuclei, which might have
been important especially in the early phase of galaxy formation and evolution.

\section{Overview of orbital decay of massive BH binaries: from kpc to sub-pc scales}

Since the seminal paper of Begelman, Blandford \& Rees (1980) it is assumed that a binary of massive BHs will evolve
under a sequence of mainly three phases before gravitational wave emission can take over and bring the binary
to coalescence: (I) the pairing phase in which dynamical friction of the galactic cores embedding the
massive black holes, which are not yet mutually bound, drives their orbital decay; (II) a second phase in which the holes
dynamically couple to form a {\it binary} and continue
to sink due to the drag caused by dynamical friction against the surrounding background of gas and stars; 
(III) a final phase in which the binary hardens via 3-body scatterings off single stars.
As we will discuss in this paper, in gaseous backgrounds there are additional aspects of the environment and nature of the drag  that complicate this simple
scheme, pointing towards the existence of more than one regime of decay at both large and small scales.
We will show, for example, how the second phase is driven not solely by conventional dynamical friction (DF) since global
disk torques are potentially at least as important.
Furthermore, in an inhomogeneous background, drag onto BHs might come from clumps, non-axisymmetric features and other sources
of gravitational torques that render the orbital decay inherently stochastic in phase II.
Additionally, the drag dominated by gas and stars will co-exist in general, with no net separation between the second
and last phase. Three-body encounters with stars may take over and
drive final coalescence in the remnants of gas-rich galaxies because of favourable orbital structure allowing to 
refill steadily the
loss cone (Khan et al. 2012), overcoming the "last parsec problem" noticed in earlier stellar 
dynamical simulations (Milosavljevic \& Merritt 2001; Berczik et al. 2006).

\begin{figure}
\begin{center}
\includegraphics[height=3.5in,width=3.5in,angle=0]{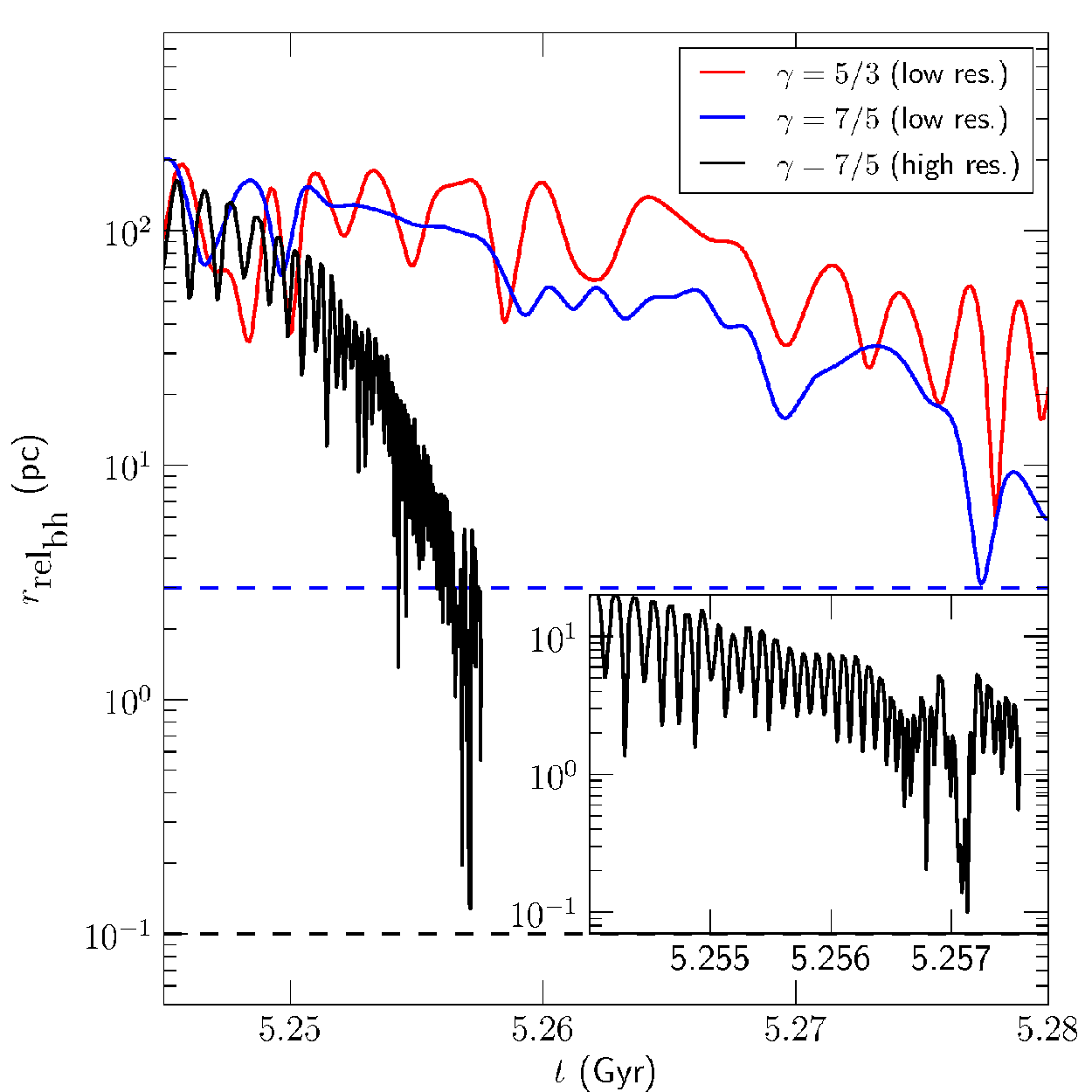}
\caption{
Evolution of the SMBH binary separation in the simulations of equal mass galaxy mergers with
equal mass massive BHs presented in Chapon et al. (2013). The simulations shown have different
resolution and polytropic index of the equation
of state. The highest resolution simulation, corresponding to the black line, reaches a spatial
resolution of $0.1$ pc while the other simulations have a spatial resolution ten times lower.
They were all carried out
with the AMR code RAMSES. The inset shows that the separation of the two massive BHs begins to
fluctuate around a fraction of a parsec, suggesting possible stalling. Courtesy of MNRAS.}
\end{center}
\end{figure}

In this paper, however, we will focus on the effect of the gas alone.
The latest results of numerical simulations suggest the following distinct regimes of BH orbital decay in
gas-rich galaxy mergers:

\begin{figure}
\begin{center}
\includegraphics[height=5in,width=5in,angle=0]{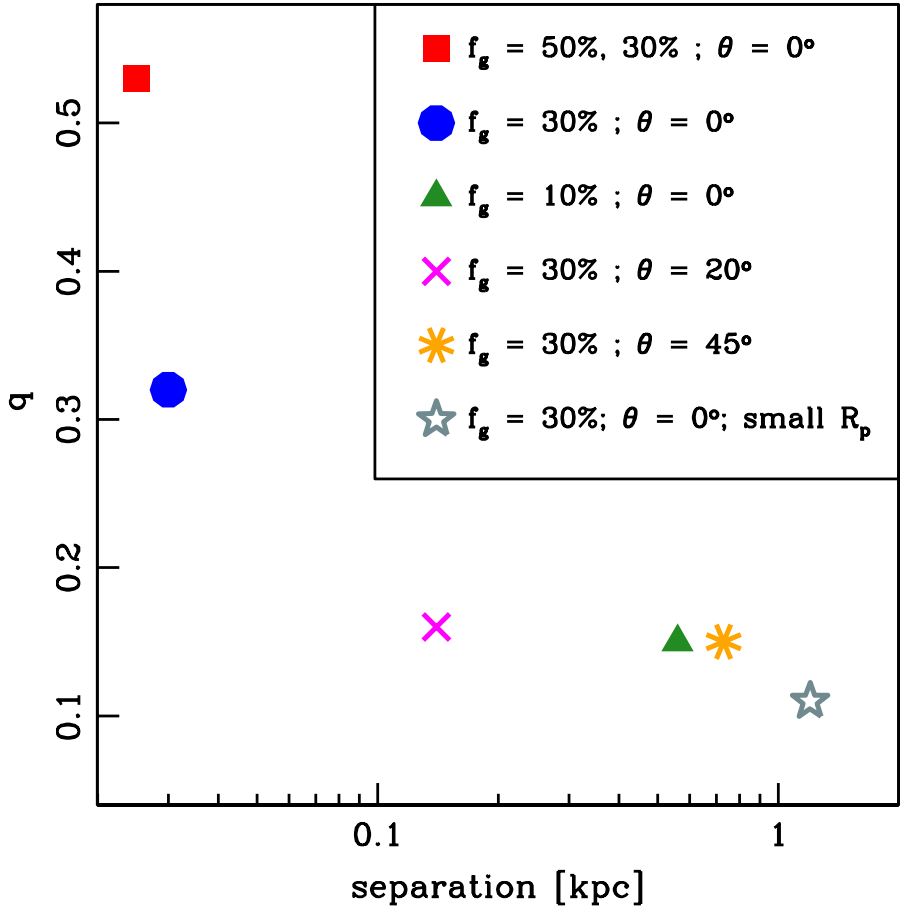}
\caption{
Outcome of the large scale evolution of massive BH pairs in minor disk galaxy mergers
(mass ratios 1:10) presented in Callegari et al. (2011).
The different symbols show the separation of the two MBHs and
  the corresponding MBH mass ratio $q$ at the time when the smaller
  of the two galaxies is disrupted by tides. The runs are labelled
  according to their initial gas fractions $f_{\rm g}$, orbital
  inclination $\theta$ and initial pericenter $R_{\rm
    p}$. Courtesy of ApJ}
\end{center}
\end{figure}

\begin{itemize}

\item {
(1) The large scale pairing of massive BHs as their host galaxy cores are still in the process of merging,
for separations from several 10 kpc to a few tens of pc, corresponding to the first phase in the
sequence of Begelman et al. (1980).}

\item {
(2) Orbital decay in a smooth circumnuclear disk driven by dynamical friction, for BH binaries with
separations of 100 pc to below
1 pc. In this case the drag results from a trailing wake excited by the moving black hole, which exerts a gravitational back-reaction
onto it. The circumnuclear disk,a few 100 pc in size, forms as a result of the merger between the two galaxies, because
gas does not dissipate completely its angular momentum via shocks and gravitational torques during the merger (Barnes 2002; Mayer et
al. 2008).}

\item{
(3) Orbital decay in a smooth circumnuclear disk driven by disk torques, analogous to Type I migration in planetary evolution
(Lin \& Papaloizou 1979; Papaloizou et al. 2007).
In this case the drag at separations below 100 pc, hence well inside the disk,
arises by torques, primarily excited at Linblad resonances and at the corotation resonance. At variance
with the dynamical friction wake, here it is mostly the gas exterior to the orbit of the BH binary that
extracts angular momentum from it.}

\item{
(4) Orbital decay in a clumpy disk: stochastic migration, also at separations of 100 pc to parsecs. This is a whole new
regime, which is expected to arise since the interstellar medium in galaxies, including galactic nuclei, is actually clumpy and
multi-phase.}

\item{
(5) Orbital decay in circumbinary disk. It occurs separations of less than a parsec (analogous to Type II 
migration in planetary evolution).

It takes place when the interaction between the target body and the rotating
gaseous background becomes nonlinear due to a large mass ratio and/or low pressure support in the
surrounding gas; disk torques from the binary
become strong enough to repel disk material away (Goldreich \& Tremaine 1980; Papaloizou et al. 1997), creating a gap and leading to slow
migration occurring roughly on the viscous timescale of a circumbinary disk (Armitage \& Natarajan 2005).
This phase might in principle lead directly
to the gravitational wave dominated regime by-passing the third phase driven by 3-body encounters present in the
Begelman et al. scenario.}

\end{itemize}

\begin{figure}
\begin{center}
\includegraphics[height=6in,width=3in,angle=270]{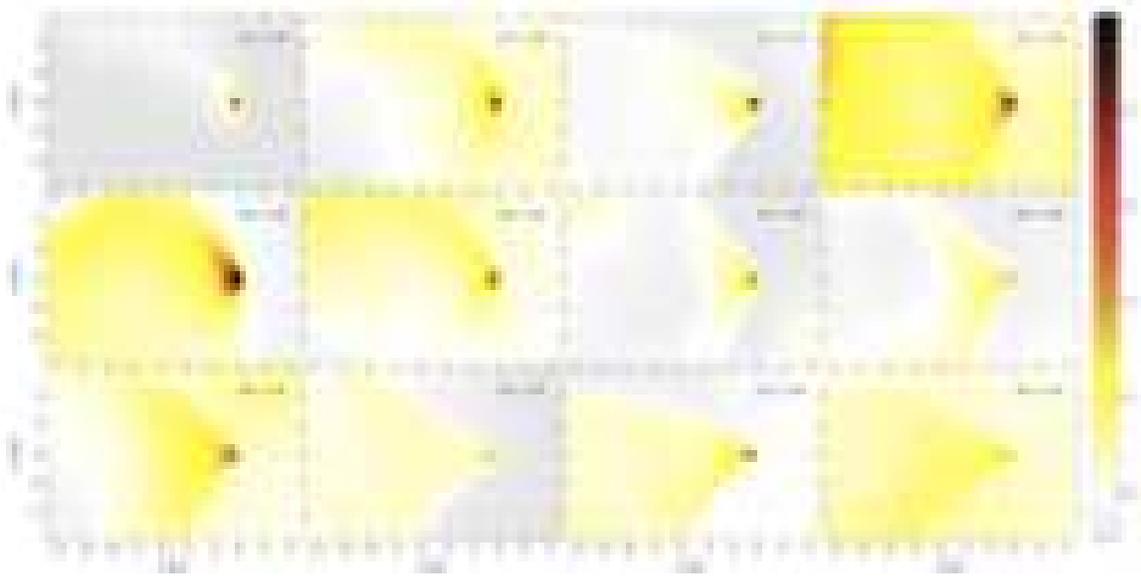}
\caption{
Color-coded density maps in the proximity of a massive BH showing the time-evolution of the 
hydrodynamical wake due of dynamical friction in one of the AMR galaxy
merger runs of Chapon et al. (2013).
The hydrodynamical wakes and Mach cones are shown for increasing values of $ {\cal M} 
=V_{BH}/c_s$.
These wakes are clearly stronger and make the dynamical friction much more efficient when the black
hole is in a transonic regime ($ {\cal M} =1.14, 1.18, 1.23$). Courtesy of MNRAS.
}
\end{center}
\end{figure}

Among these three regimes, the first two have been thoroughly studied with 3D hydrodynamical simulations, both SPH and AMR,
(5) has been studied in only a few works, (4) is beginning only now to be explored, and, finally, the distinction between (2) and (3)
has been overlooked so far. A qualitative analysis of the differences between (2) and (3) is one of the main objectives
of this paper.

Let us now proceed with a brief overview of our understanding of the orbital decay phase driven by dynamical friction.
Dynamical friction has historically been studied primarily for non-dissipative backgrounds, such as to understand
the drag exerted by a larger stellar system or dark matter halo on a smaller system or halo, such as a
dwarf galaxy satellite or globular cluster orbiting in the Milky Way halo (eg White 1978; Colpi, Mayer \& Governato 1999).
Restricted to a static, infinite, isotropic and
homogeneous collisionless medium,  the most commonly adopted approach to dynamical friction
is Chandrasekhar's formula (1943). The latter assumes that
the stars/dark-matter particles move along unperturbed trajectories
that are straight lines, thus neglecting the self-gravity present in
real backgrounds.  Despite this limitation, the resulting formula for the drag force has been often
used in a local fashion to estimate sinking times also in inhomogeneous self-gravitating
backgrounds. 

More general approaches to dynamical friction have been developed in the last two decades,
with the aim of overcoming the various limitations of Chandrasekhar's formula. Among these
we recall approaches based on global torques amplified at resonances (eg Weinberg 1989) and
the theory of linear response (TLR) (eg Bekenstein \& Maoz 1992; Colpi et al. 1999), which both attempt to capture the global response in
a finite background, taking into account at least partially
its self-gravitating nature. We refer to Colpi \& Dotti (2009)
for a thorough overview on the subject.

For illustrative purpose we can consider the example of a target body orbiting in a stellar (spherical) bulge or dark matter halo,
for which a singular isothermal sphere represents a good approximation for the density profile.
In a singular isothermal sphere with
1D velocity dispersion $\sigma$ and density profile
$\rho(r)=\sigma^2/[2\pi G r^2],$ TLR predicts a sinking time,
expressed in terms of the circularity $\epsilon$ (which is the ratio
between the angular momentum of the actual orbit relative to that of
a circular orbit of equal energy)
\begin{equation}
\tau_{df}=1.2 {r_{\rm cir}^2V_{\rm cir}
\over \ln(M_{\rm halo}/M_{BH}) GM_{BH}\epsilon^{0.4}}
\end{equation}
where $r_{\rm cir}$ and $V_{\rm cir}$ are the initial radius and
velocity of the circular orbit with the same energy of the actual
orbit, and $M_{\rm halo}$ is the mass of the dark matter and stars
within $r_{\rm cir}.$ Applied to the case of  a massive black hole,
the above equation implies that the latter
can sink at the center of the sphere within a time $\sim 10^{8}$ yr,
if released during the merger at a distance of $r_{\rm cir} \sim 100$ pc:
\begin{equation}
\tau_{df} \sim 5\times 10^8 \left (5\over {\ln {M_{\rm halo} \over M_{BH} } } \right ) 
{\left ({r_{cir}} \over 300 {\rm pc} \right )}^2
\left(V_{\rm cir}\over \sqrt{2} ~~ 100 km/s  \right)
\left( 10^6 M_{\odot} \over M_{BH} \right) \epsilon^{0.4} \rm yr
\end{equation}

Let us now consider gaseous backgrounds, still infinite and homogeneous.
In a smooth gaseous background  it is useful to consider two limits of dynamical friction onto a body moving with
velocity $V$, the supersonic regime ($V/v_s
> 1$ and the subsonic regime $V/v_s < 1$). We refer here again to the review by Colpi \& Dotti (2009).
Following Ostriker (1999) in the steady-state limit and for {\it
supersonic} motion, the drag on a massive perturber moving with
velocity $\bf V$ across a homogeneous fluid with density $\rho_{\rm
gas}$ and sound speed $c_{\rm s}$ reads
\begin{equation}
{\bf F}^{\rm gas}_{\rm DF}=-4\pi \ln\left [
{b_{\rm max} \over b_{\rm min} }
{ ({\cal M}^2-1)^{1/2}\over {\cal M}}\right ]
G^2\,{M_{bh}^2} \rho_{\rm gas}
{{\bf V}\over V^3}, \,\,\,\,\,\,\,{\rm for}\,\,\,\,\,\,\,{\cal M}>1
\end {equation}
where ${\cal M}=V/c_{\rm s}$ is the Mach number.

The deceleration
results, in this regime, by the enhanced density wake that lags behind
the perturber and that is confined in the narrow Mach cone.  Note
that  the gaseous drag is enhanced for supersonic motion (with
${\cal M}\sim 1-2.5$) compared to Chandrasekhar's formula in collisionless backgrounds (for
$\rho_*=\rho_{\rm gas}$) by a factor $\sim  2$ (compare with standard 
formula in collisionless background on eg Binney \& Tremaine 1987).

\bigskip

In the {\it subsonic} limit, instead, the drag
can be much weaker than in the collisionless case.  The
drag is exactly zero in a homogeneous infinite medium due to the
front-back symmetry of the perturbed density distribution present in
any stationary solution.  However, in a finite medium if the perturber
triggers the disturbance at time $t=0$ and moves subsonically along a
straight line, the symmetry is broken as long as $(c_{\rm s}+V)t$ is
smaller than the size of the medium, resulting in a finite drag $ {\bf
F}^{\rm gas}_{\rm DF}=-(4/3)\pi G^2\,{M_{BH}}^2 \rho_{\rm gas} {{\cal
M}^3}{\bf V}/ V^3\propto {M_{BH}}^2 \rho_{gas},{\bf V}/c_{\rm s}^3$
for ${\cal M}\ll 1.$


Consider now the case of a BH moving in a rotating background. Suppose also that the BH is moving on a circular orbit. It can be
either corotating or counter-rotating with the background. In the corotating case, which should be a natural configuration
(even in galaxy encounters with fairly high relative inclination lead to corotating BH binary configurations in nuclear
disks as a result of
large scale gravitational torques, see Callegari et al. (2011)), the BH would have not net velocity relative to the background. This implies
zero drag due to DF according to equation (3). Hence, if a drag is present in this case it has to be of a different nature. As we will
discuss in the next section, this regime can arise even if the initial orbit of the BHs is eccentric because DF itself
can circularize the orbit.

\section{Large scale pairing of massive BHs; the remarkable difference between major and minor mergers}

When the two host galaxies with their extended dark matter halos have not
completed the merger yet, the two BHs evolve as their host galaxies.
The merging timescale of halos and galaxies strongly depends on their mass ratio. Normally in
galaxy formation mergers with mass ratios larger than 1:3 are considered minor, while those with lower mass ratio
are considered major. In nearly equal mass (major) mergers the outcome is trivial; the two galaxy cores merge rapidly,
in a few Gyr, after dynamical friction onto the extended dark matter halos has eroded their relative orbital
energy and angular momentum.

Multi-scale simulations of galaxy mergers with embedded MBHs, both with SPH and AMR codes,
have shown that, after the merging of the two cores, the two BHs decay to parsecs separations, forming a tight binary
in a circumnuclear disk on timescales
as short as a few Myr for realistic assumptions on the thermodynamics of the ISM 
(Mayer et al. 2007; Chapon et al. 2013).
Additional effects due to the presence of a clumpy multi-phase ISM,
which are studied in more recent simulations (Fiacconi et al. 2013; Roskar et al. in preparation), will be described in section 5.

From a dynamical point view, a mass ratio of about 1:5 constitutes a more physically motivated boundary;
indeed for a higher mass ratio the dynamical friction timescale becomes longer
than the tidal disruption  timescale, defined as the characteristic time over which the smaller halo/galaxy loses a significant
fraction of mass due to mutual tides (Taffoni et al. 2003).
As a result, below a mass ratio of about 1:5 the smaller galaxy will be disrupted before the
two galaxy cores can merge. This has the consequence that BHs might be left wandering at kiloparsecs from the
center rather than binding into a binary, as first pointed out in Kazantzidis et al. (2005). Indeed, once
the more massive galaxy core hosting them has been dissolved by tides, their dynamical friction timescale
inside the primary galaxy can become longer than the Hubble time due to their relatively tiny mass.
Already at the halo level, the disruption timescale depends sensitively on the internal mass distribution of both the
lighter and the more massive primary halo. In particular, this translates into a dependence on halo concentration for NFW-like
halo profiles, as thoroughly addressed in Taffoni et al (2003), and could depend on the precise dark halo inner slope
if baryonic effects can flatten the cuspy density profiles typical of CDM galaxy halos as recently suggested (eg Kazantzidis et al. 
2013).

With all other things being equal, the disruption timescale can be even more severely affected by the physics of the baryonic cores.
Indeed, even if the halo is almost entirely disrupted, whether or not a dense core survives in the smaller galaxy
depends on the efficiency of gas dissipation, which acts to raise the central density and renders the core more
resilient to tides. Callegari et al. (2009;2011) found that baryonic physics determines the outcome of
minor mergers as far as the binding of BHs into a binary is concerned. The efficiency of gas dissipation depends on
the relative weight of radiative cooling vs. radiative and mechanical heating processes, such as that resulting from
feedback due to stellar irradiation and supernovae explosions. These processes also depend indirectly on the mass distribution
in galaxies and on the orbit of the galaxy encounter, which affect the strength of torques concentrating gas to the cores of
galaxies during the interaction, as well as the efficiency of star formation. This explains
the marked dependence on the pairing timescale of BHs on the orbital parameters and gas fraction of the host galaxies
found by Callegari et al. (2011), as shown shown in Figure 2. Figure 2 also shows that the conditions
in which pairing of binary MBHs is most efficient are also those in which the secondary MBH grows more by gas accretion,
which should promote a subsequent efficient hardening of the binary at smaller separations. Note that these results
hold strictly for the 1:10 mergers considered by these authors, which, however, are among the most typical merging
events in hierarchical galaxy formation (Volonteri et al. 2003).

\begin{figure}
\begin{center}
\includegraphics[height=3.5in,width=3.5in,angle=0]{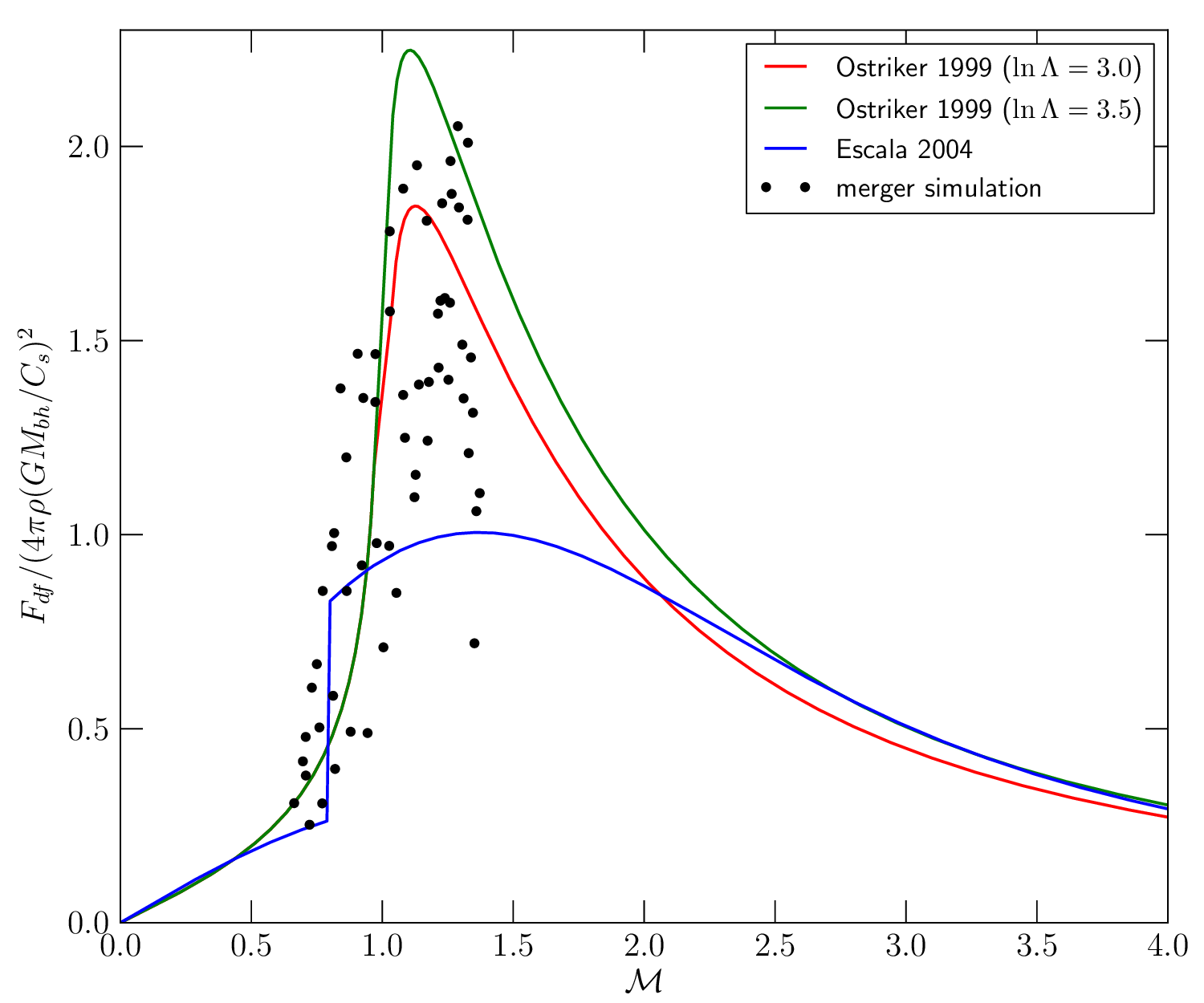}
\caption{
Comparison between measured drag force in the AMR simulations of galaxy mergers with SMBHs of Chapon et al. (2013) (black dots) and the 
predicted drag from
dynamical friction in a gaseous medium using analytical theory (Ostriker 1999) and simulations of black holes moving in spherical
gaseous backgrounds (Escala et al. 2004 - blue).
We show the renormalized dimensionsless dynamical friction force as a 
function of the Mach number (see Chapon et al. 2013). The analytical prediction is shown
for a Coulomb logarithm $\ln \Lambda = 3$ (red) and $\ln \Lambda = 3.5$ (green).
Courtesy of MNRAS.}
\end{center}
\end{figure}

Simulations are still limited by the small size of the parameter space explored in terms of 
galaxy orbits, assumptions on the internal structure of galaxies and modelling of various important processes such as star
formation, stellar feedback, and black hole accretion and feedback, which all appear to influence the pairing process.
Overall, extrapolating the orbital decay rates to smaller separations, hence assuming there is no
bottleneck at MBH binary separations below parsecs 
the simulations suggest that a binary of MBHs with initial mass ratio
of 1:10 will decay down to a separation
at which gravitational waves drive fast coalescence (less than $10^{-2}$ AU)
in a time varying from $10^8$ to $10^9$ yr. As explained in the next section, gap opening at small separations
(below $0.1$ pc) will slow down the orbital decay but the associated sinking timescale at such small separations
is still negeligible relative to that of pairing at large scales.
Note that an overall sinking timescale close to 1 Gyr for minor mergers is significantly
longer than the timescale expected for equal mass MBHs sinking in major mergers, which, as stated above, is likely below $10^7$ yr   
(Mayer et al. 2007; Chapon et al. 2013). 

\section{Orbital decay of MBH binaries in a smooth circumnuclear gas disk}

Let us first look now at the decay in a smooth gaseous circumnuclear disk.
There are obvious limitations in the analytical approach to the drag force 
presented in the previous section when the goal is to describe the orbital
evolution of MBHs in realistic galaxy mergers, in which the background
is inherently highly inhomogeneous and time-dependent, and has stars and
dark matter in addition to gas.
Nevertheless, Chapon et al. (2013),
using AMR simulations with the RAMSES code to model galaxy mergers with embedded BHs,have provided convincing
evidence that the orbital decay of BH pairs on eccentric orbits 
is well described by dynamical friction once they are embedded in the common
circumnuclear disk arising when the two galactic cores merge.
They were able to resolve with unprecedented detail the trailing wake produced by the BHs (Figure 3),
whose strength, as well as that of the associated drag force, was shown to vary with the Mach number
(defined as the ratio between the speed of the massive BH and that of the thermal sound speed)
as expected  in the analytical theory of Ostriker (1999). This is shown in Figure 4.
Chapon et al. (2013) also found that the drag was shutting off at the scale at which dynamical friction due to
the wake is expected to brake down, namely when the separation of the two holes encompasses a gas mass comparable
or lower than the sum of their own masses. The orbital decay was thus seen to stall at separations of $\sim 0.1-0.5$ pc.
Escala et al. (2006) had noticed a similar problem in their simulations of circumnuclear disks but had shown
that in some cases the decay could be restarted via local torques induced by an ellipsoidal gas deformation around
the BH binary, perhaps reminiscent of Type III migration in planetary evolution (a term used to refer to torques
acting in the co-orbital region, see Papaloizou et al. 2007).

Is this pointing to a potential {\it last parsec problem} in gaseous media, in analogy to the difficulties found
in purely stellar dynamical backgrounds when the orbital distribution of stars does not allow efficient loss cone
refilling (see Milosavljevic \& Merritt 2001; Berczik et al. 2006)?

\begin{figure}
\begin{center}
\includegraphics[height=5in,width=5in,angle=0]{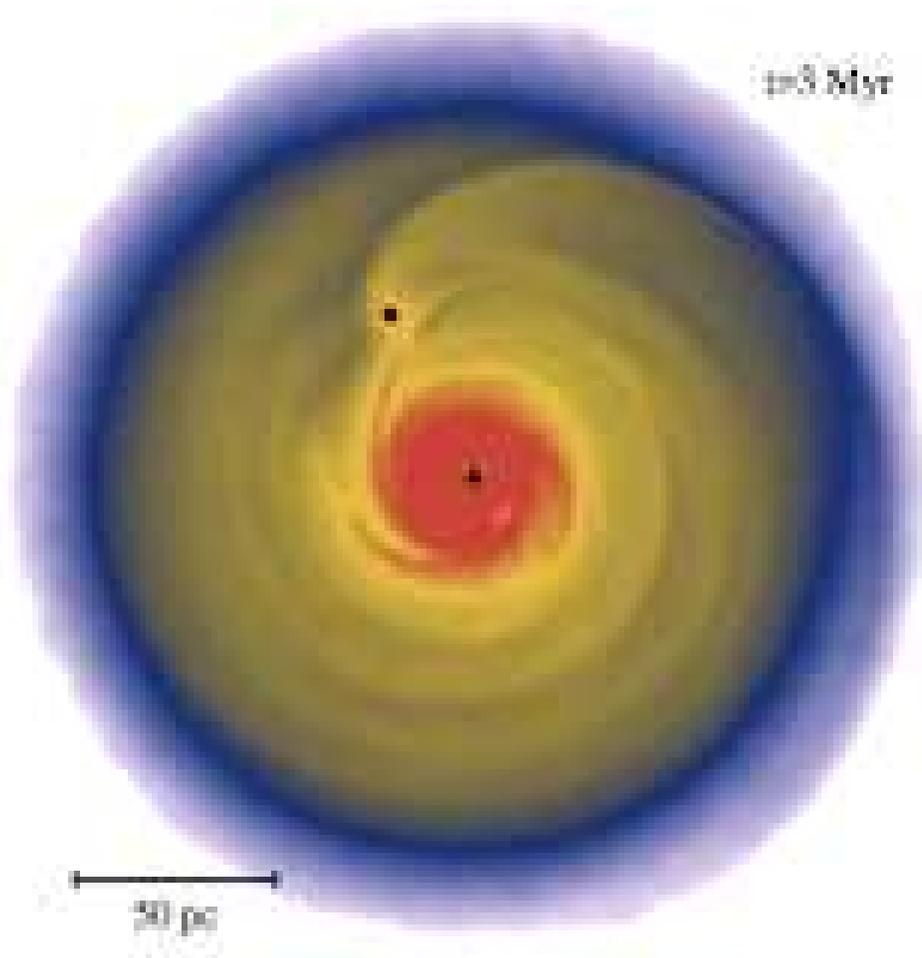}
\caption{
Color-coded density map of one of the smooth circumnuclear disk simulations (time since the beginning of the simulation, when the
secondary starts at the boundary of the disk, is indicated by the 
label) presented in Fiacconi et al. 
(2013). The spiral wave triggered by the secondary BH, a signpost of orbital decay driven by Type I-like torques, is evident.}
\end{center}
\end{figure}

The way out likely lies in the existence of the second regime in which global disk torques, rather than the local effect of the
wake, dominate the drag. Such a regime is analogous to the Type-I migration thoroughly studied in planet migration
(Papaloizou et al. 2007). In this regime a lower mass secondary black hole orbiting around the primary excites density waves
causing torques that exchange angular momentum with the hole. The spiral wave triggered by the secondary BH is clearly seen in Figure 5, which shows
a snapshot of the evolution of a BH binary in a circumnuclear disk modeled using a polytropic equation of state
(Fiacconi et al. 2013). 
In linear theory for laminar viscous disks one can see that 
the inward directed torque, which extracts angular momentum, is maximised at the outer Linblad resonance between the orbital
frequency and the spiral pattern frequency. This wins over the outward pushing torque, which peaks at the inner Lindblad resonance, as
long as there is more disk material outside the orbit of the hole rather than inside. It is thus expected that the 
negative torque will dominate only when the hole is well inside the disk. {\it The other important point is that no
saturation of the torque is expected, roughly as long as there is gas in the disk that can provide the torque exterior
to the orbit of the holes. Therefore there is no natural bottleneck of the orbital decay, contrary to the case
of the dynamical friction wake, hence no "last parsec problem" is expected in this case.}
Corotation torques can also provide
drag, although this depends on the details of the flow (Papaloizou et al. 2007). 
In Figure 6, which shows the evolution of the angular momentum of the secondary BH for a set of simulations such
as that shown in Figure 5, it is evident that there is a first phase of slow decay and then a second phase of faster
decay. The first phase is well described by dynamical friction while the onset of the second phase coincides with
the appearance of a strong triggered spiral density wave, as shown in  Figure 5.

Assuming linear torque theory, Type I decay rate scales inversely with the mass of the decaying BH as in the case of dynamical friction.
An expression for the Type-I migration timescale, which works pretty well for relatively small planets, being
borne out of linear theory, hence corresponding to light  secondary 
black holes in our case (with masses $< 10^{-2}$ that of the disk) is due to Tanaka (2002):

\begin{equation}
\tau_{migr} = {(2.7 + 1.1 \beta)}^{-1} {{{M_1}^2} \over {M_2 \Sigma a^2}} h^2 {\Omega}^{-1}
\end{equation}

where $M_1$ is the mass of the primary black hole, $M_2$ the mass of the secondary black hole,
$a$ is the orbital separation of the two black holes, $\Sigma$ the disk surface density, $h$ its scale height and $\Omega$
its angular velocity.
The equation is derived assuming laminar, non-self gravitating disks, and includes
3D effects and corotation torques, for a disk with surface density profile $\Sigma \sim r^{-\beta}$, scale
height $h$ and orbital frequency $\Omega$. We can assume $\beta = 1$, as in the convenient
Mestel disk model adopted in many simulations (see also section 6).
For secondary black holes with masses up to $M_2 = 10^6 M_{\odot}$ this equation should be applicable given the large masses of circumnuclear disks adopted in the
simulations ($M_d \sim > 5 \times 10^8 M_{\odot}$). For reasonable values of the parameters ($M_1 = 10^7 M_{\odot}$, $\Sigma = M_d/2 \pi {R_d}^2$, with $M_d = 5 
\times 10^8 M_{\odot}, R_d = 100$ pc, $a = 10$ pc, $h = 5$ pc, $\Omega = v_c/ a$, with the circular velocity $v_c$ being a constant
in a Mestel disk, $v_c = 100$ km/s) the result is $\tau_{migr} \sim 1$ Myr, in substantial agreement with
the numerical results (see Figure 6).
Recent simulations of massive planets embedded in self-gravitating, turbulent  protoplanetary 
disks,in which gap opening is difficult even at large planetary masses, show that the
mean migration rate, although non-monotonic, still occurs
roughly on the Type-I migration timescale (Baruteau et al. 2011).

Note that this regime of orbital decay was not observed in the published simulations of galaxy mergers  (Mayer et al. 2007;
Chapon et al. 2013), which instead seem to be well understood based on the effect of the local trailing wake of
dynamical friction all the way down to the smallest BH separation reached at the center of the disk.
The fundamental reason behind the difference with respect  to the results found in the simulations of circumnuclear disks is not clear yet. However, the
fact is that in the circumnuclear disk simulations the Type-I migration regime is normally preceded by a phase
of orbital circularization induced by dynamical friction, and in cases in which the orbit is already nearly
circular in the initial conditions the type-I phase indeed starts earlier (Figure 5). Instead, in the merger simulations the orbit
of the black hole binary remains eccentric and the transition to the Type-I regime never occurs (Mayer et al. 2007). 
The role of circularization might be just coincidental but certainly needs to be clarified since it appears to mark
the transition between orbital decay driven by the dynamical friction wake and by global disk torques.
The other important ingredient to enter an efficient Type-I phase
might be the presence of a relatively cold, weakly turbulent disk in which the large scale spiral modes 
induced by the sinking BHs, which torque 
them as a back-reaction, are not weakened by large pressure gradients or turbulence. The lack of favourable thermodynamical
conditions might partially explain why the type-I like regime is not seen the multi-scale merger simulations
of Mayer et al. (2007) and Chapon et al. (2013), which have rather hot and turbulent disks.

The conditions under which orbit circularization operates effectively are still unclear. Circularization is important
not only to understand the transition to the type-I regime but also because the merger rate of BHs once they enter
the gravitational wave regime is a strong function of orbital eccentricity. 
There is evidence that 
circularization requires a rotating
background, be it gaseous or stellar. Indeed, detailed studies of orbital eccentricity evolution in non-rotating stellar and 
gaseous backgrounds have never observed circularization even  after a large number of orbits; among these the studies of orbital
decay of satellite galaxies and/or dark matter halos (eg Colpi et al. 1999). Circularization has
been observed recently in rotating spherical stellar backgrounds (Sesana et al. 2011), and had been demonstrated earlier in rotating circumnuclear
disks of stars and gas (eg Dotti et al. 2007). 
Dotti et al. (2007) provide the following qualitative explanation of why circularization occurs. They relate it to the position of the
wake at apocenter versus pericenter; they find that at apocenter the wake is in front of the black hole, causing an increase in
its tangential velocity because disk material rotates at a speed faster than  that of the black hole, causing a positive torque,
while at pericenter the wake 
is behind the black hole and causes a decrease of its velocity. The net result is an increase of the ratio between tangential and
total velocity of the black hole over an orbit, which effectively leads to a lower eccentricity. 
This would not happen without rotation, since in this case the wake is always behind the black hole. The net circularization
will be dependent on the rotation curve of the disk since that will determine the magnitude of the effect at apocenter.
On the other end, rotation in the background is clearly not a sufficient condition for circularization. Indeed, as
we already noticed in the the multi-scale
simulations of galaxy mergers with embedded BHs circularization is not observed (Mayer et al 2007; Chapon et al. 2013).
The difference might be that in these mergers the resulting circumnuclear disk is more massive and self-gravitating, 
with self-sustained spiral waves
that could force the eccentricity to remain large, as it has been shown to be 
possible for migrating planets in asymmetric
disks (Papaloizou \& Larwood 2000). 


Star formation would reduce the density of the gaseous disk, increasing both the dynamical friction and the type-I migration timescales.
However, it has been shown that, for comparable disk mass distribution and thickness, BHs sink at similar rates in a mainly gaseous or mainly stellar disk (Dotti
et al. 2007), suggesting that neglecting the effect of star formation in this entire discussion of the sinking regime is a reasonable approximation.

Even if it enters a fast Type-I phase the decay may still stall if the BHs are able to open a gap, as it is well known in 
planetary evolution. This happens when the gap-opening gravitational torque exerted by the rotating binary onto the surrounding gaseous medium
prevails over the gap-closing viscous torque associated with turbulent/viscous diffusion in the circumbinary gas.
A useful form of the gap opening condition was derived by Escala  et al. (2005), under the assumption that the viscous/turbulent
speed characterizing the magnitude of the gap-closing torque is equivalent to the effective sound speed in the nuclear disk,
that a clear gap has a size $\Delta r \sim 3 h$, where $h$ is the disk scale height set by the balance between pressure
and gravitational forces (the criterion was calibrated on simulations' results), and that the BH obeys the observed $M_{BH}-\sigma$ relation
between black hole mass $M_{BH}$ and central stellar velocity dispersion $\sigma$ (this being proportional to the total mass within
the orbit of the BH binary) :

\begin{equation}
{M_{BH} \ge {(h/pc)}^2 \times 7.2 \times 10^5 M_{\odot}}
\end{equation}

Assuming a mean disk thickness of 30-40 pc, corresponding the effective sound speed of 50-60 km/s, consistent
with the gas velocity dispersions in local circumnuclear disks of interacting or merging
galaxies, such as ULIRGs and LIRGs (Downes \& Solomon 1998; Medling et al. 2013)
one finds that the mass at which BHs are expected to open a gap is quite large, of order
$10^9 M_{\odot}$. This neglects the complication of a multi-phase ISM, in which the cold molecular gas phase, distributed
in the midlpane of the circumnuclear disk, might have
a shorter scale height, of order $5-10$ pc. Still, even for a massive BH binary confined to the  
thin cold molecular gas layer a fairly massive pair of BHs will be needed to open
a gap, with a total mass $M_{BH} > 10^7 M_{\odot}$ based on equation (5). However, in the latter case
the lower limit is small enough that BHs in massive early-type spirals as well as in typical elliptical galaxies would be able to open a gap, 
making the regime relevant to the general evolution of BH binaries.

\begin{figure}
\begin{center}
\includegraphics[height=3in,width=3in,angle=0]{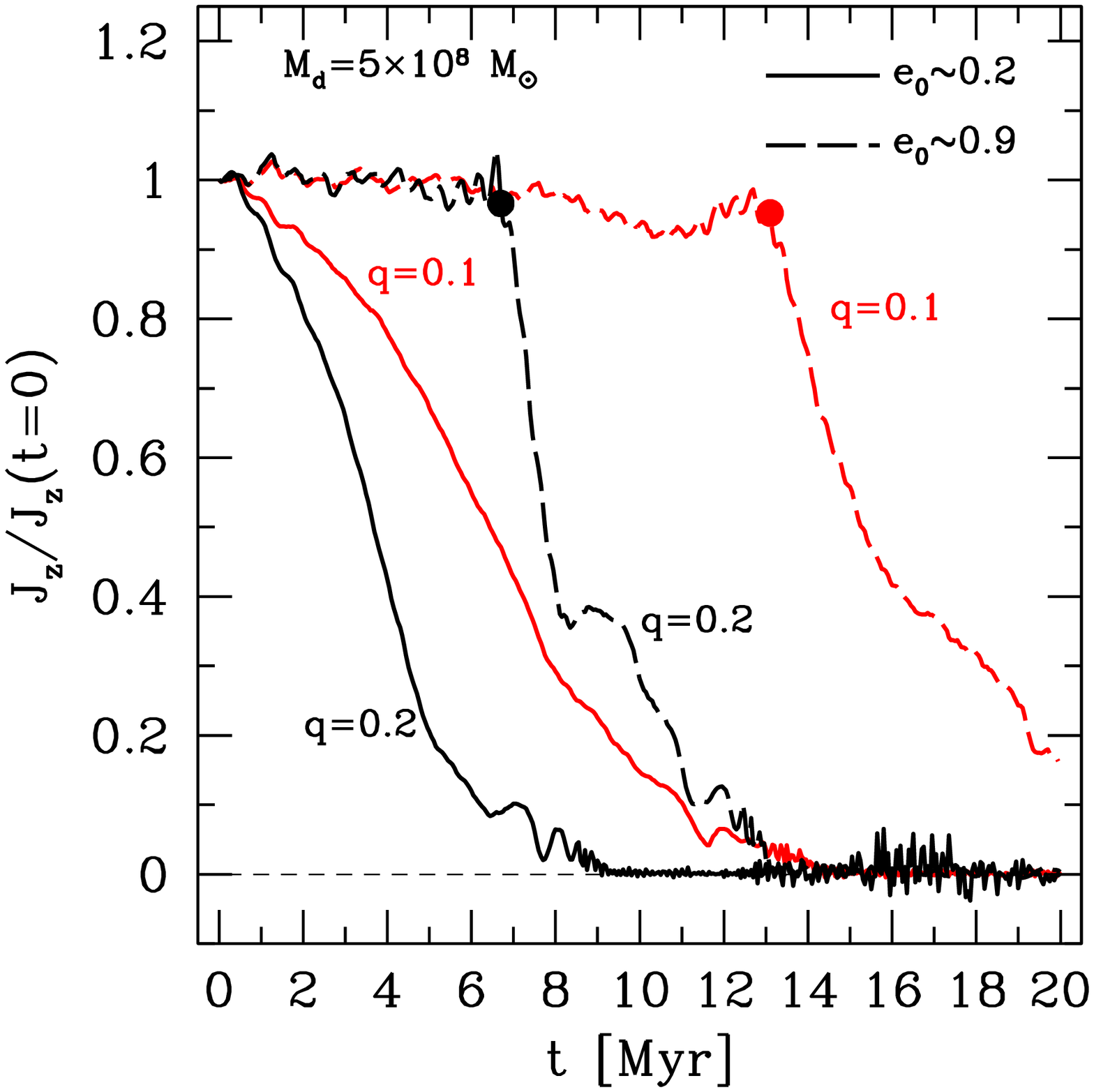}
\includegraphics[height=3in,width=3in,angle=0]{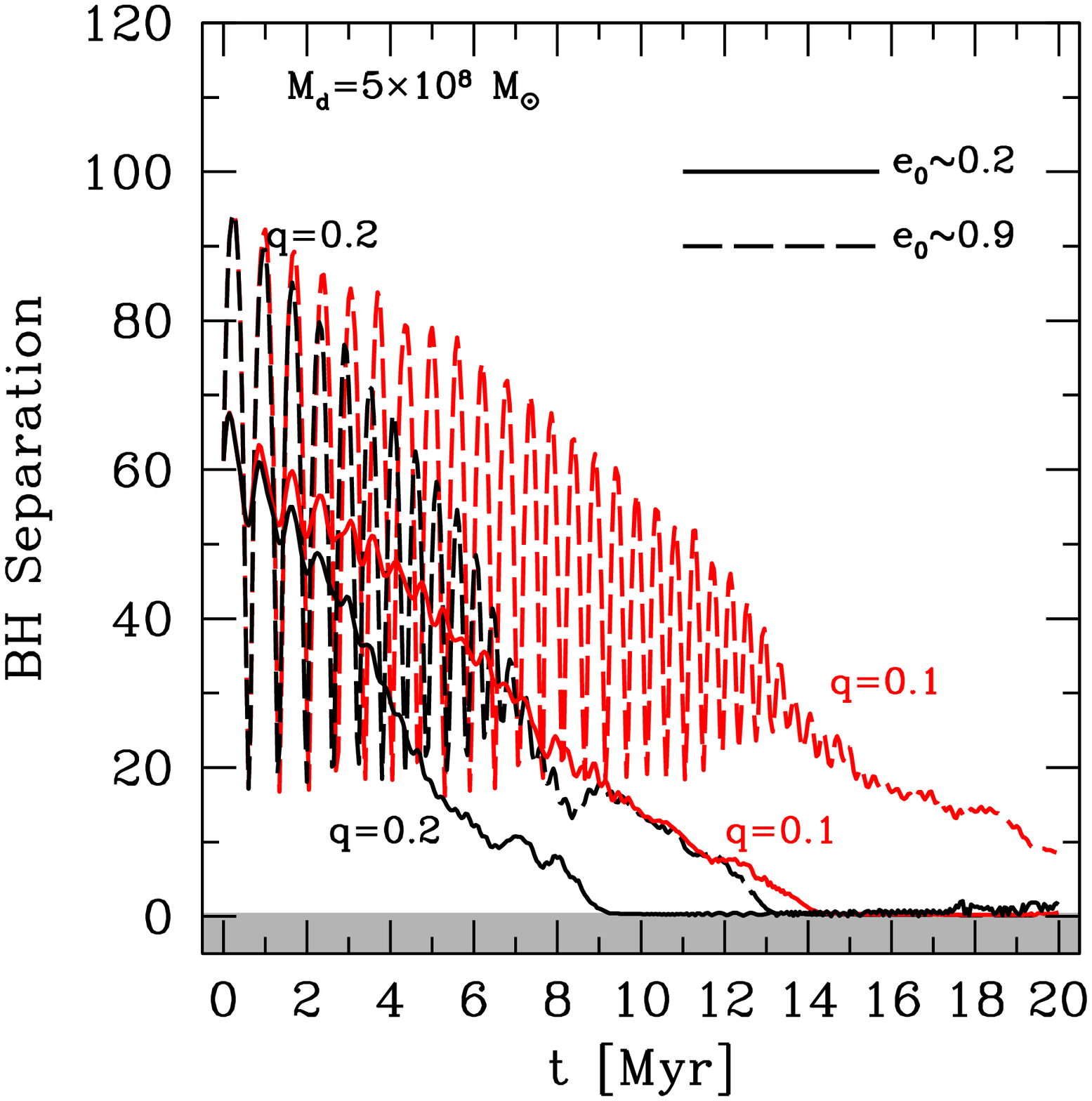}
\caption{
Time evolution of the angular momentum (left) and separation of two massive BHs (right) with two 
different mass ratio $q$ on two different orbits with eccentricity $e$ as indicated in the
labels. The massive BHs evolve in a smooth circumnuclear disk (see Fiacconi et al. 2013).
The orbital eccentricity diminishes as a result of dynamical friction, as can be seen
by the fact that the angular momentum evolves little while the separation decreases in the
first part of the evolution. After the orbit circularizes significantly the last part of the 
decay becomes faster and is driven by disk torques resulting from the back-reaction of the
spiral waves triggered by the secondary BH, as shown in the Figure 5.}
\end{center}
\end{figure}

The possibility of gap opening has motivated studies which model 
the evolution of a massive BH binary at small separations ($a < 0.1$ pc) within a gap embedded in a common circumbinary 
disk (Armitage \& Natarajan 2005; Cuadra et al. 2009).
The orbital decay in this configuration can still proceed on a timescale proportional to the viscous timescale of the sub-pc 
scale circumbinary accretion disk with which the binary exchanges angular momentum. Following Armitage \& Natarajan (2005),
the binary orbit will shrink at a rate:

\begin{equation}
{da/dt \propto - \alpha{(h/r)}^2 q_d \Omega a}
\end{equation}

where $q_d = \pi a^2 \Sigma / M_2$, $a$ being the semi-major axis of the binary orbit, $\Sigma$ the disk surface density,
$\Omega$ the angular frequency of the gas, $h$ the disk thickness and and $\alpha$ the viscosity, with $\alpha$ in the range
$10^{-4} - 0.1$ depending on the mechanisms that can generate viscosity, such as the magneto-rotational instability or
gravitational (Toomre) instability. For $\alpha \sim 10^{-3}$
equation (6) leads to a merger timescale of order $10^7$ yr according to simple 1D numerical newtonian models which simply add the phase of 
decay via gravitational wave emission to obtain the total merger timescale (Armitage \& Natarajan 2005). Note that such timescale
is comparable or longer than the orbital decay timescale required for BHs to sink from 100 pc to pc scale separations
in galaxy merger simulations or circumnuclear disk simulations ($1-10$ Myr).

Global 3D simulations exploring the circumbinary disk stage across many orbits are needed to consolidate the estimate of the decay 
timescale but so far they have probed only a relatively small number of disk orbits, over which the binary separation decreases 
by only a few percent (Cuadra et al. 2009; Roedig et al. 2012). 
These simulations incorporate self-gravity and show that, if the circumbinary disk is massive, the effective viscosity that
governs transport of angular momentum is dominated by gravitational torques acting between the binary and the disk perturbed
by the binary. Torques from material accreting onto the black hole play an important role too, and in some cases are crucial
to ensure that overall a negative torque is felt by the binary, promoting its shrinking (Roedig et al. 2012).
However, they do not include effects associated with
magnetic fields that might be important in the dense, hot gas sitting close to the potential well of the black holes.
Indeed in the circumbinary accretion disc the gas becomes gradually warmer and more ionized towards smaller radii, so 
that faster angular momentum transport can arise by MRI or, more in general, via magnetic stresses
(Menou \& Quataert 2001). If this is the case a deep gap would never be cleared as MHD stresses push more matter inside the gap, allowing 
the binary to couple more strongly with the surrounding gas and hence be torqued more effectively. This is essentially what has been 
found in recent General-Relativistic MHD simulations of circumbinary disks, where the binary shrinkage occurs nearly 3 times faster than 
in hydrodynamical simulations (Shi et al. 2012).

\section{Orbital decay of a MBH binary in a clumpy disk; stochastic decay}

In a gravitoturbulent nuclear disk, such as that which should arise in presence of self-gravity, cooling and heating/stirring via feedback
mechanisms (Wada \& Norman 2001; Agertz et al. 2009), the medium is rapidly filled with cold and dense gaseous with sizes comparable with
molecular clouds (1-10 pc). A simple way to produce a gravitoturbulent state, without including the various heating and cooling
source terms directly, is to introduce a phenomenological dissipative term in the internal energy equation (Fiacconi et al. 2013).
This dissipation term effectively represents the net cooling rate in the system. If it is large enough massive, gravitationally
bound clumps are produced from rapid fragmentation.
The resulting gravoturbulent state is an extreme version of  a clumpy medium since the resulting clumps are compact and
tightly bound while molecular clouds are fluffy objects, often unbound, 
subject to internal feedback from stellar radiation, stellar winds and supernovae,
Yet these experiments are useful to understand
and quantify the role of a clumpy medium in the context of the orbital decay of BHs
(Fiacconi et al. 2013). The main consequence is that orbital decay becomes stochastic.
BHs can either be slowed down or accelerate their inward migration, and even be ejected temporarily from the disk plane
depending on how they exchange energy and angular momentum in gravitational scatterings with clumps.
Massive clumps, corresponding to the scale size/mass of Giant Molecular Clouds (GMCs), namely of order $10^6 M_{\odot}$ or larger,
are mostly responsible for scattering holes outside the disk plane.

Once outside the disk plane the hole can sink back down into the gas disk owing to dynamical friction from the stars and dark
matter. Most of the contribution to the mass, hence to the drag, comes from the stars. The drag in this regime is
much smaller than in the disk since the average density of the background is much smaller and dynamical friction
is also slightly less effective than in gaseous backgrounds (see also Escala et al. 2005).
For a range of bulge
parameters and stellar density profiles (from Plummer to Sersic-like distribution with Sersic index in the range $1-2$, to represent
from classical bulges to pseudobulges), Fiacconi et al. (2013) find that the decay timescale back to the disk varies from
20 Myr to nearly $10^8$ yr. Once back in the disk the holes can shrink to sub-pc scale separations in less than 1 Myr
even in clumpy disks, confirming previous results with smooth media in which the BH was never leaving the disk plane
(Mayer et al. 2007; Dotti et al. 2007; Chapon et al. 2013). Hence it is the time spent inside the stellar-dominated
background that introduces a bottleneck in the decay process.

Note that at $z > 2$ massive disk galaxies are observed to be clumpy at kpc scales, with clumps of masses approaching even
$10^7-10^8 M_{\odot}$, hence much larger 
than the mass of present-day GMCs. While it is not clear yet whether or not such clumps are 
long-lived bound condensations produced by gravitational instability in a turbulent ISM as the clumps in the
simulations of Fiacconi et al. (2013), scattering of massive BHs out of the plane of the
nuclear disk should thus occur even more at high redshift.
Since at $z > 1$ the detection rate by future gravitational wave experiments should be fairly high, it will be important
to clarify the quantitative role of clumpiness in a large set of simulations.
Clumps can also aid the orbital decay of MBHs by binding with one of them, thus increasing the effective mass subject to
friction and/or disk torques, as found by Fiacconi et al. (2013).

We estimated the BH mass threshold below which these stochastic effects due to a clumpy medium are going to be important. This
is given by:

\begin{equation}
\mathcal{M}_{\bullet}\sim7\times10^{7}\,\bigg(\frac{N_{\rm cl}}{4}\bigg)^{1/4}\bigg(\frac{\eta}{0.4} \bigg)\bigg(\frac{M_{\rm g}}{10^{9}\;{\rm M_{\odot}}}\bigg)\,{\rm M_{\odot}},
\end{equation}

where $\eta$ is the gas fraction of the circumnuclear disk, $M_g$ its (gas) mass and $N_{cl}$ the number of very massive
clumps of mass comparable to that of the sinking BH.
From the above equation we see that, at low-z, when clump masses are at most as large as those of GMCs
($\sim 10^6 M_{\odot}$), black holes with up to a few $10^7 M_{\odot}$ can be affected by 
gravitational scattering with the
clumps, while at high-z, when gas clumps possibly produced by gravitational instability can reach masses as large 
as $10^8 M_{\odot}$ according to observations of galaxies at high-z (Genzel et al. 2011), black holes as large as
$\sim 10^9 M_{\odot}$ can be affected, essentially spanning nearly whole mass spectrum of massive BHs in present-day galaxies.

A further step forward in the modelling of a clumpy medium can be achieved by replacing the single dissipative term with a sub-grid model
for a multi-phase ISM, including optically thin radiative cooling, absorption and scattering of radiation at high
optical depths via a phenomenological temperature-density relation calibrated with radiative transfer calculations,
star formation and supernovae feedback using the blastwave approach successfully employed in galaxy formation
simulations with the GASOLINE SPH code (see eg Mayer 2010). Multi-scale galaxy merger simulations
reaching $0.1$ pc resolution with the aid of particle splitting have been carried out with all the latter ingredients
and will be presented in a forthcoming paper (Roskar et al. 2013). An example of such computationally demanding
state-of-the-art simulations, which comprise several millions gas and star particles in the nuclear disks emerging from the merger,
timesteps as small as a few hundred years and a spatial resolution of $0.1$ pc, is shown in Figure 7.
In these simulations a gravoturbulent medium arises naturally, with
a broad mass spectrum of cloud/clumps, from a few thousand solar masses to $> 10^6 M_{\odot}$. The effect of clump-BH scattering
is confirmed to delay the orbital decay significantly due to ejections out of the disk plane,
bringing it up to nearly $10^8$ yr in some instances, in substantial agreement with the results of Fiacconi et al. (2013).

\begin{figure}
\begin{center}
\includegraphics[height=3in,width=3in,angle=0]{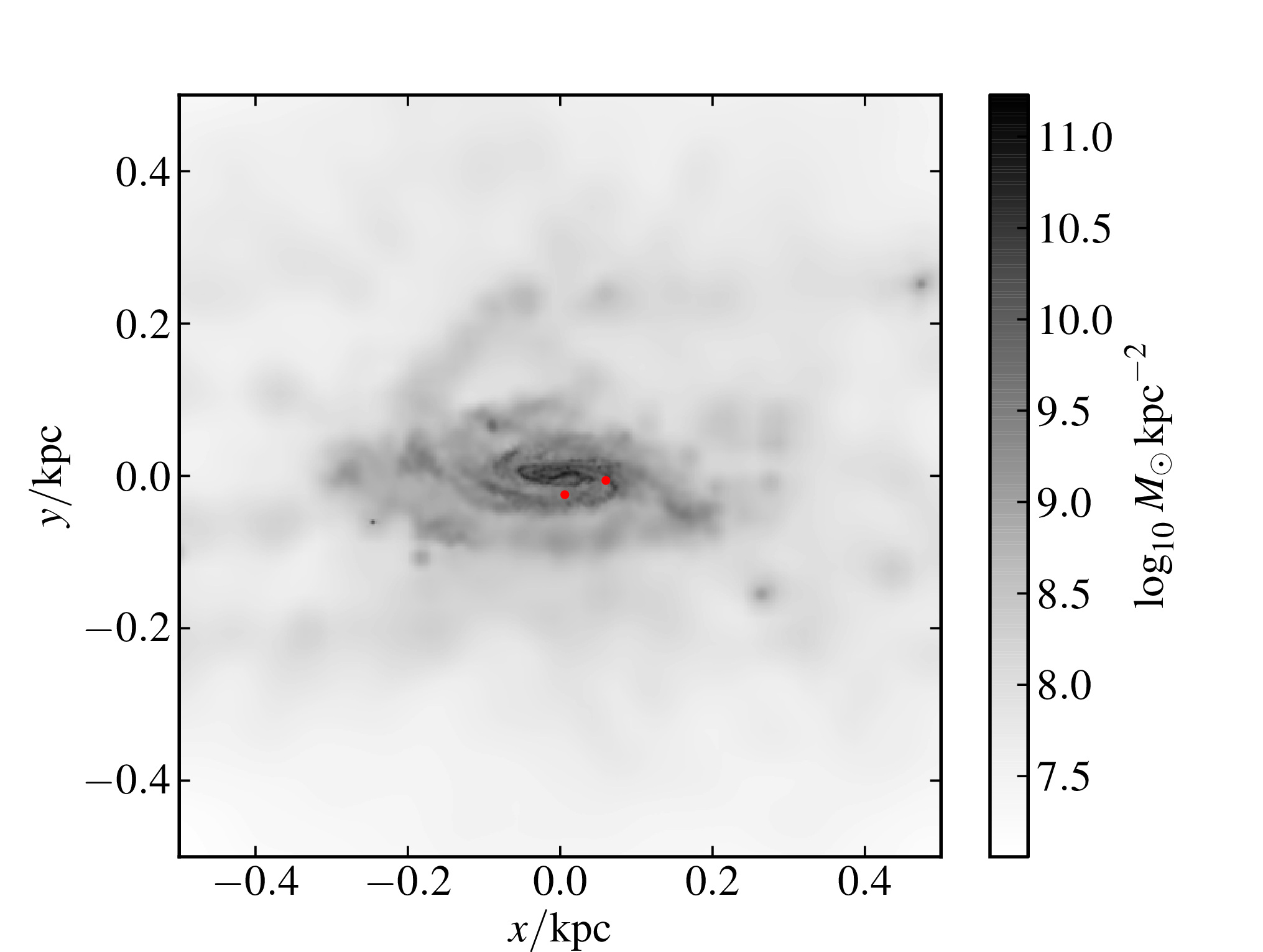}
\includegraphics[height=3in,width=3in,angle=0]{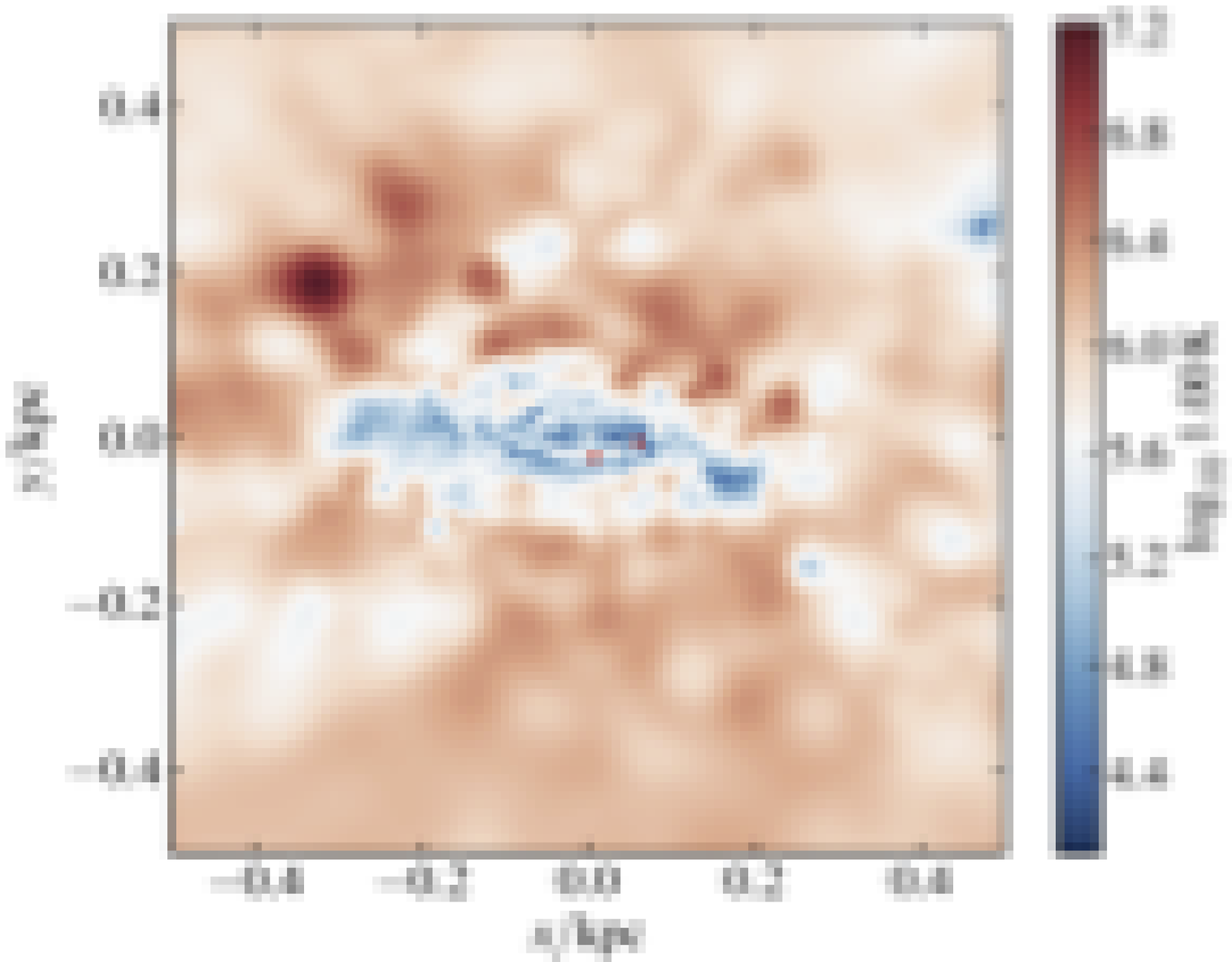}
\caption{
Projected density map (left) and temperature map (right) showing two inspirling massive BHs in 
the clumpy, multi-phase circumnuclear disk forming at the end of a gas-rich galaxy merger in the 
multi-phase SPH simulations of Roskar et al. (2013, to be submitted). The snapshot corresponds to 
the time after the BHs have  just returned to the plane of the disk after one of them had been 
ejected to the interaction with a massive clump. The resolution is a few thousand solar masses 
and $0.1$ pc, owing to particle splitting.}
\end{center}
\end{figure}



\section{Interplay between orbital decay of massive BHs and inflow-driven massive BH seed formation}

Among massive seed BH formation mechanisms many rely on multi-scale gas inflows, either triggered
by non-axisymmetric instabilities in marginally unstable protogalaxies, such as bars-within-bars 
(Begelman et al. 2006;
Lodato \& Natarajan 2006; Regan \& Haenhelt 2009;
Choi et al. 2013; Latif et al. 2013 ) or by merger-driven inflows in collisions between gas-rich massive galaxies at
high redshift (Mayer et al. 2010; Bonoli et al. 2013). Mayer et al. (2008) noticed that in circumnuclear disks
that are strongly non-axisymmetric, such as those that could feed a massive BH seed im the aforementioned
models, the orbital decay of a massive BH binary stalls at a separation of parsecs (the simulations
had a resolution as high as $0.1$ pc). They argued that the reason was that in 
such disks the gas inflow is so prominent and fast to overtake the BH binary in its journey to the center;
essentially the BHs undergo only weak friction
when they are still at several parsecs from the center
because gas is funnelled to the center very rapidly
,reducing the ambient density considerably and hence the strength of the drag.
The implication is that, in order
for the binary to shrink to sub-pc separations and coalesce in a gaseous environment, the circumnuclear disk has to be relatively stable
so that the inflow does not overtake the holes. If this is true 
the epochs of massive BH seed formation and efficient massive BH coalescence are likely distinct, since they require different
conditions in the galactic nucleus. These conclusions, however, were based on
the result of a single very hi-res simulation with a fixed mass of the two BHs (which were of equal mass, $\sim 3.6 \times
10^6 M_{\odot}$). 

We can explore further the overall relevance of this result, as well as if it is physically grounded, by building a 
simple analytical model. We want to compare the
torque exerted by dynamical friction onto a massive BH binary in a circumnuclear disk with the viscous torque
the causes the gas inflow. Transport of angular momentum via spiral waves, bars or other kinds of non-axisymmetric
instabilities in self-gravitating disks can indeed be described based on an effective viscosity to zeroth order
(Lin \& Pringle 1987; Lodato \& Natarajan 2006; Mayer et al. 2010).
As we mentioned above, type-I migration torques are also involved in  the orbital decay,
but they dominate only at small separations between the holes  when the orbit of the binary has circularized, hence we can rightly
consider the dynamical friction dominated phase as the main bottleneck.
We will adopt a simple alpha-disk model (Shakura \& Sunyaev 1973), with a mass distribution described by
a Mestel disk, widely used in simulations (eg Dotti et al. 2007; Fiacconi et al. 2013). The latter is a convenient
choice to keep the equations simple.
A Mestel disk has indeed the nice feature that it has a constant circular velocity $v_c$
and its surface density is given by $\Sigma = \Sigma_0*(R_0/R) = M_d(2\pi {R_0}^2)R_0/R$, where $M_d$ is the
disk mass, $\Sigma_0$ and $R_0$ normalization factors for the surface density and the radius, respectively.
This model
strictly applies to a razor-thin disk but we will make the assumption that this is a good approximation
for a disk in which $h/R \sim 0.05$ (having a finite scale height will be useful in order to be able to
include the thermodynamics via consideration of the disk vertical structure).
The magnitude of the viscous torque between two adjacent disk annuli at $R$ can be written as:

\begin{equation}
\Gamma_{visc}= 2 \pi \nu \Sigma {R^3} d \Omega / dR
\end{equation}

where $\Omega$ is the angular velocity and  we set the viscosity $\nu = \alpha c_s h$, where $\alpha \le 1$ in a standard 
Shakura-Sunyaev alpha disk, $c_s$ is the sound speed and $h$ the effective disk scale height.
This has to be compared with the dynamical friction torque acting on a black hole binary of mass $M_{BH}$ and moving
at speed $v_{BH}$.  We will consider the supersonic case only ($V_{BH} > c_s$) since this is the regime
in which orbital decay is more efficient (see section 2)
and also the more relevant one for massive BHs delivered in the typical
conditions of galaxy mergers (Chapon et al. 2013, see also Figure 3). In this case we can use equation (3) for the dynamical
friction force $F_{DF}$ so that the magnitude of the corresponding torque 
$\Gamma_{df} = |F_{DF} R|$ is given by:

\begin{equation}
\Gamma_{df} = {{4 \pi G {M_{BH}}^2 \Sigma \ln {\Lambda} R} \over  {h V_{BH}^2}}
\end{equation}

where we have replaced the usual volume density $\rho_{gas}$ with the surface density $\Sigma$ of the Mestel disk 
, assuming for simplicity a constant disk scale height $h$ and simply that $\rho_{gas} = \Sigma/h$.
The condition that orbital decay is efficient is equivalent to the condition that the inflow does not overtake the two holes
as they decay, and can be written as $\Gamma_{df} > \Gamma_{visc}$. Combining (8) and (9) yields 
the following condition on the magnitude of $\alpha$:

\begin{equation}
\alpha < 2 {G {M_{BH}}^2 \ln {\Lambda}  \over c_s {V_{BH}}^2 v_c h^2}
\end{equation}

where ln$\Lambda$ is given by the expression in equation (3), i..e ln$\Lambda \sim  3$ for 
$b_{max}/b_{min} \sim 10$ (which
reflects the ratio between disk size and distance of the holes from the center where stalling was observed in
the simulations of Mayer et al. (2008)) and Mach numbers in the range $1-2.5$ as typical
for BHs embedded in circumnuclear disks on orbits with a range of eccentricities (see Chapon et al. 2013),
Furthermore, we adopt $V_{BH} \sim 200$ km/s, assuming that the
circumnuclear disk is hosted in a relatively massive galaxy (note that the velocity will actually increase as the holes
sink further in the potential well of the disk).
We note that the surface density 
cancels out in the above equation
but the assumption of a Mestel disk is still important since it imposes a simple law for the profile
of $\Omega(r)$ given the constancy of $v_c$, so that we could replace $|d \Omega /dR| = |v_c / R^2|$ in equation (8).
{\it For $M_{BH} = 10^8 M_{\odot}$, $c_s$ = 60
km/s, $h = 5$ pc, $R= 10$ pc, $v_c = V_{BH}$, we obtain $\alpha < 0.1$}. 

Let us now examine this result in context. 
For non-turbulent self-gravitating
disks an effective viscosity $\sim 0.1$ is often interpreted as a threshold between fragmentation and transport
of angular momentum via gravitational torques, as long
as one assumes that local linear perturbation theory applies to describe gravitational instability, so that one can
adopt the local Toomre stability criterion for axisymmetric waves as a global empirical instability criterion for
generic modes. In this case the condition $\alpha < 0.1$ corresponds to $Q < 1$, which is the standard
threshold  for fragmentation in a thin disk (Lodato \& Rice 2004;2005).
Hence within this simple framework the fact that we find a critical $\alpha$ close to such threshold has
a simple and sensible physical meaning; it implies that
dynamical friction will be effective as long as there is a stable background, which in this
case is the circumnuclear disk.

If $\alpha$ grows beyond the threshold, however, the inflow
will overtake the orbital decay only temporarily since with $Q < 1$ the disk is expected to fragment, turning gas into stars at a high
rate, so that the decay might be restarted later as the disk becomes smooth again one most of the gas has been
converted into stars (but this time the drag will be due to friction against a stellar background).
However, 3D global disk simulations, performed at various scales and in various contexts, from protogalaxies to protoplanetary
disks, show that disks can achieve a gravoturbulent state, with star formation and gas inflows co-existing, and 
an "effective" $\alpha \sim 1$ (eg Escala
2007; Agertz et al. 2009). Such highly gravoturbulent state would seem to be 
problematic for the orbital decay 
of binary BHs based on equation (10). 
If anything, the dynamical friction timescale should lengthen considerably. The multi-phase mergers 
currently in progess (Roskar et al. 2013) account such higher physical complexity and will shed light
on these issues.

If we start by fixing the viscosity parameter, setting $\alpha = 0.1$, we can rearrange equation (10) to express a
condition on the minimum black hole mass needed for dynamical friction to be effective in presence of a strong inflow
. This, for the same parameters just adopted, it turns out to be $M_{BH} > 10^8 M_{\odot}$, which is larger than the
mass of the BH in the most massive spirals.  
As we said, since a realistic disk can generate a strong inflow in  a gravoturbulent state with an even greater effective viscosity
in such cases our same argument would imply that the inflow 
will overtake the binary for black holes even larger than $10^8 M_{\odot}$.

In order to gain further insight we can examine the case of very strong inflows, such as those postulated 
to be capable of forming massive BH seeds
at high-z during galaxy mergers (Mayer et al. 2010), as well as more moderate gas inflows, such as those necessary to supply the
gas to nuclear starbursts in local ULIRGs and feed AGNs. We will derive the expected effective $\alpha$ and compare with the condition
expressed by equation (10). We can relate the inflow rate with the effective viscosity in the circumnuclear disk via the following 
equation, which expresses the maximum inflow rate in a steady-state condition since it essentially
neglects any angular momentum or non-radial motion in the flow:

\begin{equation}
dM/dt = 2 \alpha {c_s}^3 / G
\end{equation}

In the simulations of Mayer et al. (2010), inflow rates as high as $10^4 M_{\odot}$/yr have ben measured. We can further
adopt an effective sound
speed in the range 60-100 km/s, hence a "warm" disk (which favours high inflows at comparatively smaller viscosities relative to the 
case of a "cold" disk) in order to provide a conservative estimate of the corresponding value of $\alpha$ for the inflow.
With $100$ km/s and $\alpha=0.1$ we obtain $dM/dt 
\sim 100 M_{\odot}$/yr. While this is a larger number, it is about two orders of magnitude lower than what
measured in inner tens of pc of the circumnuclear disks that undergo a central collapse into a supermassive cloud,
a possible precursor to a massive BH seed (Mayer et al. 2010; Bonoli et al. 2013). This implies effective viscosities
of order unity at least. Adopting a weaker constraint, namely the minimum inflow rate necessary to assemble
a supermassive star (SMS) of $> 10^6 M_{\odot}$, which is $> 1 M_{\odot}$/yr
(see Begelman 2010), one would still need $\alpha > 0.1$.

In any case, the conclusion is that the conditions required for massive BH seed formation via direct gas collapse 
are orthogonal to those necessary for efficient BH binary decay, which explains the findings of Mayer et al. (2008).
Likewise, circumnuclear disks have to enter a sub-critical phase with $\alpha < 0.1$, in 
which central runaway gas collapse cannot take place, in order to allow efficient coalescence of BH binaries.
This might well be the natural condition some time {\it after} a massive BH seed has formed and the host 
galaxy merges with another galaxy containing another massive BH (in direct gas collapse models the
inflow is a really quick event, operating on timescale less than $10^7$ yr, see eg Mayer et al. 2010).
Of course, even in super-critical nuclei the binary BHs might still merge, albeit on a longer timescale, as 
the stellar background  can still cause drag via dynamical friction and 3-body encounters with stars.
Finally, in presence of more moderate gas inflows such as those needed to feed "normal" QSOs and AGNs (i.e. not the
the brightest QSOs at hi-z that might require feeding at a rate $> 10 M_{\odot}$/yr) once a massive
BH is already in place there is no problem for the orbital decay.
Indeed, inflow rates just below those at the critical viscosity threshold are still prominent, being
large enough to provide enough gas supply to the accretion disk to grow already
existing SMBHs at the typical rates suggested by observations of  low redshift AGNs. These are typically
occurring at a few percent of the Eddington limit, with accretion rates in the range  $10^{-2} - 0.1 M_{\odot}$/yr (Raimundo et al. 2012),
which would correspond to $\alpha < 10^{-3}$ for the lowest accretion rates, posing thus no problem for the orbital decay of
black holes with masses even as small as $M_{BH} \sim 4 \times 10^6 M_{\odot}$ (comparable to the
mass of the black hole in the Milky Way), as it can be seen by combining equation (10) and (11).

\section{Summary and Conclusions}

We have reviewed the various phases of the orbital decay of pairs of massive BHs, from galactic scale separations to sub-pc scale separation. We have neglected the 
presence of the stellar background in our analysis, and we have not considered the effect of BH accretion and its energy feedback. The concurrent effect of the stellar
background and of black hole accretion has been explored only sporadically in the context of the orbital
decay of massive BH binaries, although BH accretion in the Bondi-Hoyle spherical approximation and thermal feedback have been included in recent
works (Callegari et al. 2011; Van Wassenhoeve et al. 2012), while hybrid calculations following the gaseous background at larger scales and 3-body stellar
encounters at smaller BH separations have been 
recently pioneered by Khan et al. (2012). A detailed study of the effect of BH accretion and 
feedback in high-resolution simulations of circumnuclear disks, adopting more realistic 
accretion models rather than the Bondi-Hoyle approximation, is currently under way.

We have presented evidence that, at least in some cases, the decay of the secondary BH undergoes a transition from a regime dominated by dynamical 
fiction to one governed by global
disk torques, which resembles type I migration in planetary evolution. This latter regime seems to become dominant after the orbit of 
the secondary BH has been
circularized by dynamical friction,
and leads to an acceleration of the decay, which can bring the binary to harden to $0.1$ pc separations in only a few Myr.
Most importantly, since negative disk torques continue to act as long as there is disk material outside the orbit of the BH, a condition 
that is increasingly
fulfilled as the BH sinks towards the center, this regime has no saturation, allowing in principle to bring the secondary BH 
to the gravitational wave
dominated regime in $< 10^7$ yr. On the contrary, recent simulations of galaxy mergers in which the Type I-like regime is not achieved
show that the BHs might stall at separations just below a parsec, since the local dynamical friction wake vanishes at 
such small separations. Understanding better the conditions that activate such efficient orbital
decay regime driven by disk torques will be of paramount importance in the future.
If gap opening intervenes below pc separations, in conditions that might be satisfied only by the most massive black holes, $> 10^8 M_{\odot}$,
the decay might slow down, with timescales of order $10^8$ yr to reach the gravitational wave dominated regime.

The results just recalled hold strictly in an homogeneous circumnuclear disk. In a clumpy, inhomogeneous disk, which is a more faithful model of the multi-phase medium
with cold molecular clouds expected in real circumnuclear disks, scattering by gas clumps/clouds and other effects render the decay stochastic, in some
cases slowing down the hardening to pc scales and below by more than an order of magnitude. In the latter case the binary will reach $0.1$ pc
separations in $\sim 10^8$ yr (Fiacconi et al. 2013; Roskar et al., in prep.).

All these timescales, either in a homogeneous or clumpy disk, are reasonably short -- even in the worst case the two black holes should merge on
a timescale comfortably below $10^9$ yr. Instead, the real bottleneck might be at larger scales, precisely larger than the size of circumnuclear disk,
about a few hundred parsecs. Indeed, while for major mergers there are no particular issues, minor mergers simulations show that, depending on the orbit 
and combined effect of gas cooling, star formation,
feedback processes etc., the two black holes might reach the nucleus of the remnant over a time from as short as a few $10^8$ yr to as long as a Gyrs,
or even not reach it at all, becoming wandering black holes (Callegari et al. 2009;2011). 
Since minor mergers are the preferred mode of galaxy assembly in a CDM universe, future
work will have to explore a much larger parameter space with simulations, with the aim at building statistical semi-analytical models that can describe
properly the large scale pairing of massive BHs as a function of the various relevant parameters. In fact, at the moment assessing the probability
distribution of the BH pairing timescales at scales $> 100$ pc might be the most pressing issue in order to evaluate how likely is that black holes
merge when their host galaxies merge.

Finally, we have begun to unveal an intriguing interplay between efficient orbital decay and efficient formation of massive BH seeds by direct gas collapse,
showing that the conditions demanded by one process essentially exclude the other one. This is because prominent gas inflows evacuate the background
that is responsible for the drag of the holes. Implicitly, this line of reasoning assumes that the stellar background does not provide an efficient
drag. If true, the existence of these two mutually exclusive regimes points to distinct phases of massive BH evolution, an early one in which massive BH seed
formation is very likely owing to prominent gas inflows in hi-z mergers among the most massive galaxies at $z \sim 8-10$ or efficient gas accretion in 
protogalaxies in very high sigma peaks, such as the first halos that can cool via atomic hydrogen, and a later one in which central gas inflows fade away
but BH binaries can merge more efficiently at the center of galactic nuclei. It is tempting to speculate on the effect that this would have on the gravitational
wave event rates from massive BH mergers as a function of redshift. For example, at early epochs  BH binaries might be 
held back in their race to the
center due to the inflows overtaking them, at late epochs mergers of massive BHs become rare because the merger rate of galaxies and 
halos as well as the gas content of galaxies decrease
significantly, while at intermediate epochs BH binaries could find ideal conditions to sink efficiently to the center as inflows
are moderate enough but merger rates and gas content are still high enough.
This and other intriguing aspects of the co-evolution of galaxies and massive BH binaries will have to be investigated
with semi-analytical models following simultaneously BH seed formation via direct gas collapse and the processes that lead to 
BH mergers in realistic backgrounds of both gas and stars.

\bigskip

\ack

\bigskip

We thank all the people that have had a long term collaboration with
the author on the topic of co-evolution of galaxies and SMBHs, and who contributed to many of the results discussed in this
review; Simone Callegari, Monica Colpi, Stelios Kazantzidis, Massimo
Dotti, Davide Fiacconi, Damien Chapon, Andres Escala, Piero Madau, Rok Roskar, 
Romain Teyssier, James Wadsley, Jillian Bellovary and Tom Quinn.
We also thank the Kavli Institute for Theoretical Physics at UC Santa Barbara for
hospitality during completion of this work while the program "A Universe of Black Holes"
was taking place. The program generated stimulating thoughts and discussions that have
been crucial to develop this work.

\end{document}